\documentclass[aps,prc,amsmath,10pt,twocolumn,superscriptaddress,showpacs]{revtex4-1}
\usepackage[dvipdfmx]{graphicx}
\usepackage{color}
\usepackage{bm}
\usepackage{multirow}

\begin{document}

\title{Matrix- and tensor-based recommender systems for the discovery of currently unknown inorganic compounds}
\author{Atsuto \surname{Seko}}
\email{seko@cms.mtl.kyoto-u.ac.jp}
\affiliation{Department of Materials Science and Engineering, Kyoto University, Kyoto 606-8501, Japan}
\affiliation{Center for Elements Strategy Initiative for Structure Materials (ESISM), Kyoto University, Kyoto 606-8501, Japan}
\affiliation{JST, PRESTO, Kawaguchi 332-0012, Japan}
\affiliation{Center for Materials Research by Information Integration, National Institute for Materials Science, Tsukuba 305-0047, Japan}
\author{Hiroyuki \surname{Hayashi}}
\affiliation{Department of Materials Science and Engineering, Kyoto University, Kyoto 606-8501, Japan}
\affiliation{Center for Materials Research by Information Integration, National Institute for Materials Science, Tsukuba 305-0047, Japan}
\author{Hisashi \surname{Kashima}}
\affiliation{Department of Intelligence Science and Technology, Kyoto University, Kyoto 606-8501, Japan}
\author{Isao \surname{Tanaka}}
\affiliation{Department of Materials Science and Engineering, Kyoto University, Kyoto 606-8501, Japan}
\affiliation{Center for Elements Strategy Initiative for Structure Materials (ESISM), Kyoto University, Kyoto 606-8501, Japan}
\affiliation{Center for Materials Research by Information Integration, National Institute for Materials Science, Tsukuba 305-0047, Japan}
\affiliation{Nanostructures Research Laboratory, Japan Fine Ceramics Center, Nagoya 456-8587, Japan}

\date{\today}

\begin{abstract}
Chemically relevant compositions (CRCs) and atomic arrangements of inorganic compounds have been collected as inorganic crystal structure databases.
Machine-learning is a unique approach to search for currently unknown CRCs from vast candidates.
Herein we propose matrix- and tensor-based recommender system approaches to predict currently unknown CRCs from database entries of CRCs.
Firstly, the performance of the recommender system approaches to discover currently unknown CRCs is examined.
A Tucker decomposition recommender system shows the best discovery rate of CRCs as the majority of the top 100 recommended ternary and quaternary compositions correspond to CRCs.
Secondly, systematic density functional theory (DFT) calculations are performed to investigate the phase stability of the recommended currently unknown CRCs.
The phase stability of the 27 compositions reveals that 23 currently unknown compounds are stable.
These results indicate that the recommender system has great potential to accelerate the discovery of new compounds.
\end{abstract}

\maketitle
\section{Introduction}
A reliable and quantitative prediction of chemically relevant compositions (CRCs) where stable crystals are formed is highly demanded because the synthesis of new inorganic compounds is an important target of physicists, chemists and material scientists.
Density functional theory (DFT) calculations have played a central role in such predictions.
However, exhaustive DFT calculations without prior knowledge of the crystal structures are expensive even for a given composition.
Recent developments and the popularization of machine learning (ML) have facilitated the prediction of currently unknown CRCs.
For example, a ML model with respect to only the composition, which was estimated from a DFT database of the formation energies, predicts currently unknown CRCs\cite{PhysRevB.93.085142}.

Another ML-based approach to search for currently unknown CRCs uses inorganic crystal structure databases such as the Inorganic Crystal Structure Database (ICSD)\cite{bergerhoff1987crystal}. 
This approach estimates the CRC probability for a candidate composition using a given compositional similarity.
A procedure to estimate the CRC probability uses the compositional similarity on the basis of the entries in the database itself\cite{hautier2010finding,hautier2010data}.
A different procedure adopts the compositional descriptors defined by a set of well-known elemental representations\cite{PhysRevB.95.144110} as prior knowledge\cite{seko-recommender1}.
Compositional descriptors are useful to avoid the so-called ``cold-start'' problem, which occurs when few known CRCs are available\cite{seko-recommender1}.
The cold-start problem prevents the prediction of CRCs due to the lack of known CRCs.

These approaches based on inorganic crystal structure databases can be referred to as ``recommender systems'' for the discovery of currently unknown CRCs.
Recommender systems\cite{resnick1997recommender,aggarwal2016recommender} developed in the ML community have become increasingly popular in a variety of scientific and non-scientific areas.
In the field of commerce, recommender systems suggest items to a user.
Such a recommender system predicts the rating or preference for an item from an existing dataset comprised of the users' history such as items purchased and numerical ratings given to items to generate a recommendation for the user. 

Simple matrix and tensor-based recommender system approaches can be adopted to discover currently unknown CRCs.
These approaches show a robust performance of recommendations for a wide variety of datasets in the ML community\cite{sarwar2000application,frolov2017tensor,symeonidis2016matrix}.
In this study, we build recommender systems to discover currently unknown CRCs using matrix and tensor-based approaches, which are descriptor-free approaches.
We employ different approaches: non-negative matrix factorization (NMF), singular value decomposition (SVD), canonical polyadic (CP) decomposition, and Tucker decomposition.
NMF and SVD are matrix-based recommender systems, while CP and Tucker decompositions are tensor-based recommender systems.
Their performances to discover currently unknown CRCs are compared.
Then using a list of compositions recommended by the best recommender system, we predict the phase stability of compounds using DFT calculations, and examine the performance of discovering currently unknown CRCs using the predicted phase stability.

\section{Methodology}

\begin{figure}[tbp]
\begin{center}
\includegraphics[width=0.8\linewidth,clip]{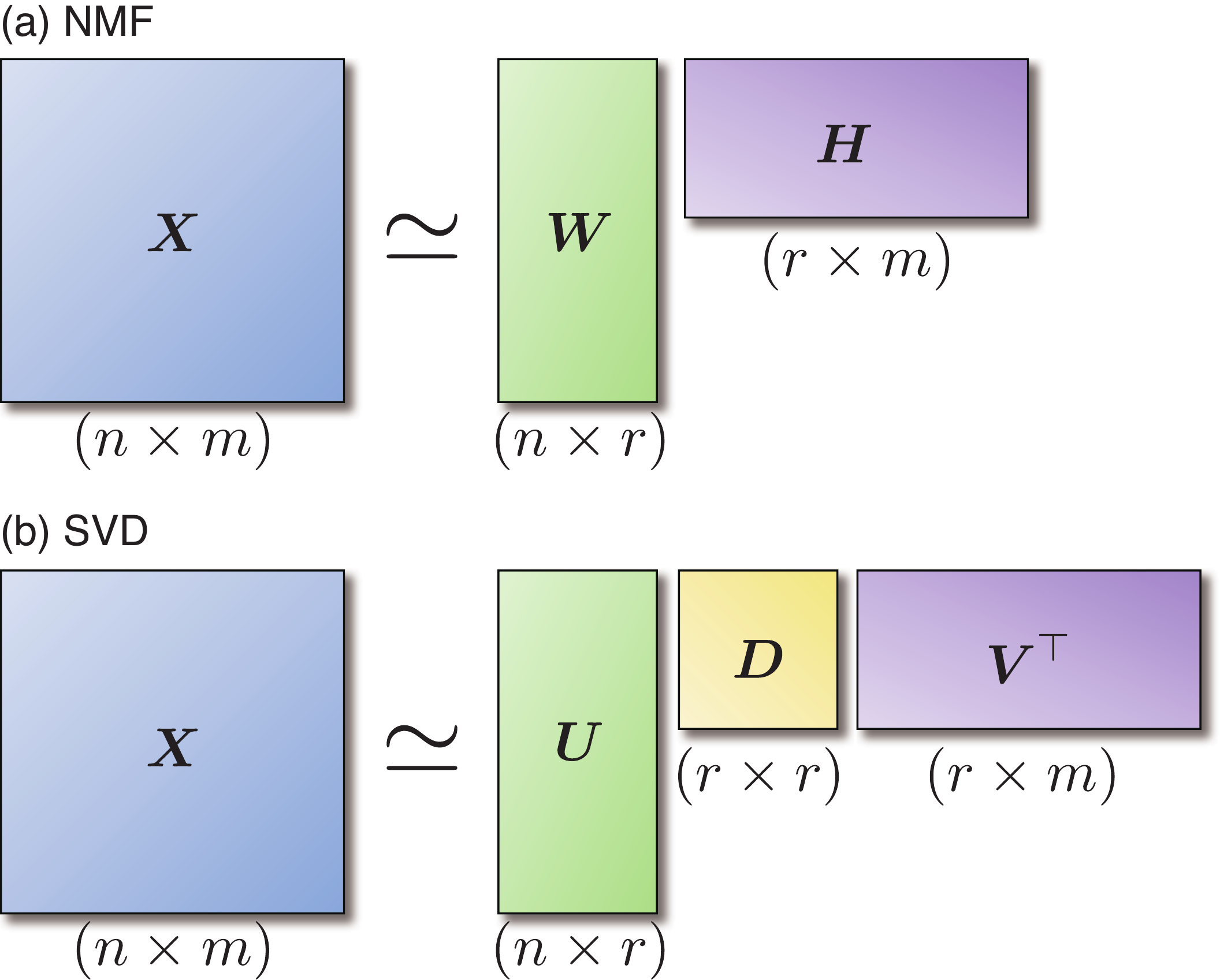} 
\caption{
Schematic illustration of the approximation of the rating matrix by (a) NMF and (b) SVD.
Dimensions of factorized matrices are also shown.
}
\label{recommend:Matrix_decomposition}
\end{center}
\end{figure}

\subsection{Matrix factorization}

A historical dataset used to build a recommender system is simply described as a form of an incomplete matrix or tensor with missing values, called a ``rating matrix'' or a ``rating tensor''.
The underlying assumption of many recommender system approaches is that the complete rating matrix or tensor has a low-ranked structure.
Using this assumption, the ratings of missing elements are predicted as an approximated rating matrix or tensor with a given reduced rank.

The nonnegative matrix factorization (NMF) factorizes nonnegative rating matrix $\bm{X}$ into two nonnegative matrices $\bm{W}$ and $\bm{H}$ with given rank $r$. 
$\bm{W}$ and $\bm{H}$ do not have negative elements.
Figure \ref{recommend:Matrix_decomposition} (a) illustrates the behavior of NMF.
The original $(n, m)$-dimensional matrix $\bm{X}$ is approximated as
\begin{equation}
\bm{X} \simeq \bm{W}\bm{H}
\end{equation}
using $(n, r)$-dimensional matrix $\bm{W}$ and $(r, m)$-dimensional matrix $\bm{H}$ with a given rank $r$.
Each of the $r$ dimensions corresponds to a distinctive rating trend.

Another matrix factorization approach is the singular value decomposition (SVD).
Figure \ref{recommend:Matrix_decomposition} (b) illustrates the behavior of SVD.
SVD factorizes original rating matrix $\bm{X}$ into three matrices.
The rating matrix is approximated as
\begin{equation}
\bm{X} \simeq \bm{U}\bm{D}\bm{V}^\top
\end{equation}
using $(n, r)$-dimensional matrix $\bm{U}$, $r$-dimensional diagonal matrix $\bm{D}$ and $(m,r)$-dimensional matrix $\bm{V}$.
The diagonal matrix contains only $r$ largest singular values. 
Hence, each of the $r$ dimensions corresponds to a distinctive rating trend as well as the NMF.
The approximated rating matrix is obtained using the algorithms implemented in \textsc {scikit-learn}\cite{scikit-learn}.

\begin{figure}[tbp]
\begin{center}
\includegraphics[width=0.9\linewidth,clip]{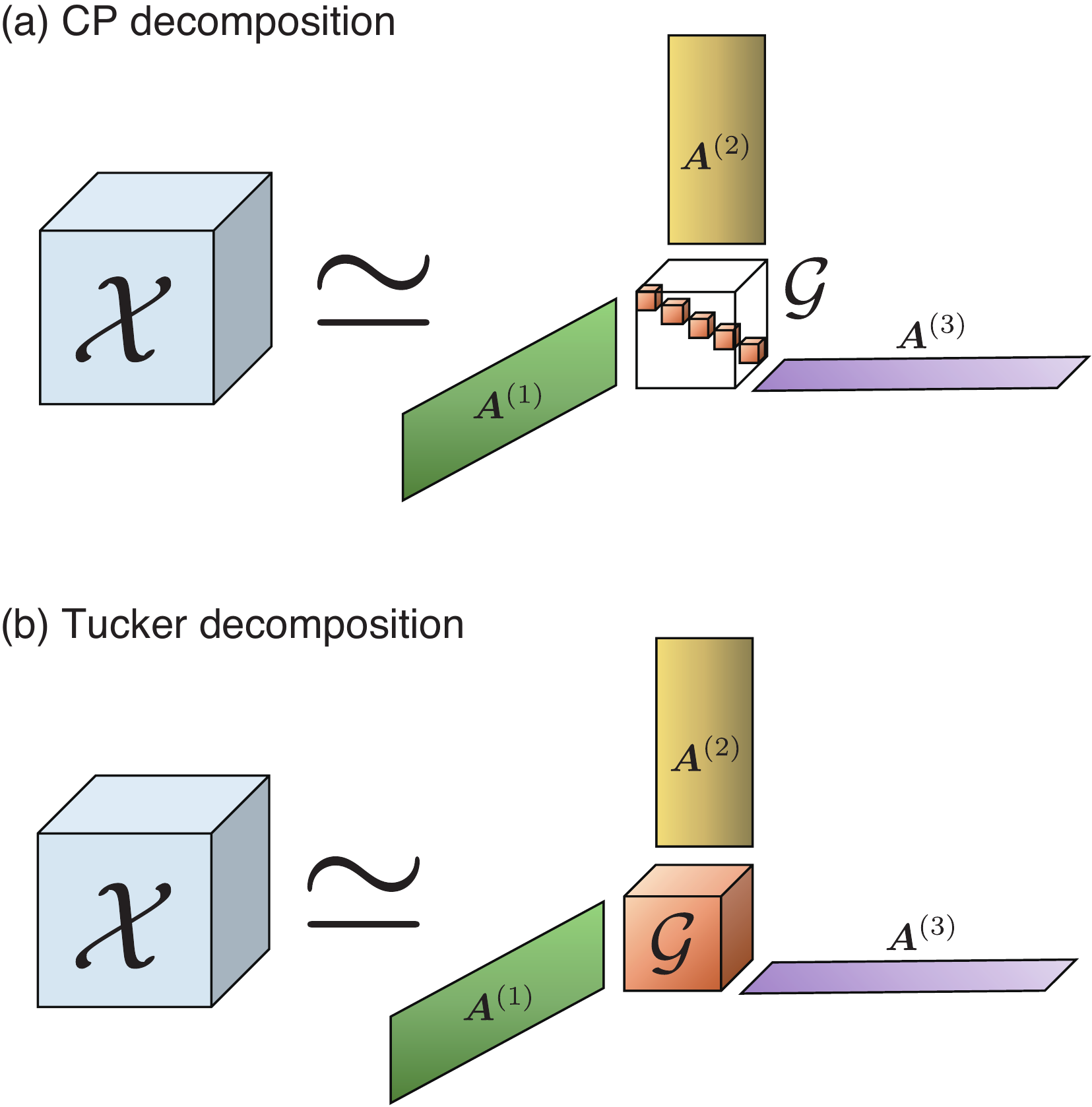} 
\caption{
Schematic illustration of (a) CP and (b) Tucker decomposition.
}
\label{recommend:Tensor_decomposition}
\end{center}
\end{figure}

\subsection{Tensor factorization}
Matrix factorization approaches require that the original data is transformed into a matrix form when the rating depends on three or more factors.
On the other hand, tensor factorization approaches can include information of all factors naturally as a form of the rating tensor.
A tensor factorization technique is the tensor rank decomposition or canonical polyadic (CP) decomposition\cite{hitchcock1927expression}.
Figure \ref{recommend:Tensor_decomposition} (a) illustrates the behavior of CP decomposition.
CP decomposition factorizes rating tensor $\mathcal{X}$ into a set of matrices and a super diagonal core tensor.
Third-order rating tensor $\mathcal{X}$ is approximated using three matrices $\bm{A}^{(1)}$, $\bm{A}^{(2)}$, and $\bm{A}^{(3)}$ as
\begin{equation}
\mathcal{X} \simeq \mathcal{G} \times_1 \bm{A}^{(1)} \times_2 \bm{A}^{(2)} \times_3 \bm{A}^{(3)}, 
\label{recommend:eqn-tensor}
\end{equation}
where $\mathcal{G}$ denotes the core tensor.

A more flexible tensor factorization is the Tucker decomposition\cite{tucker1966some}, which is also known as higher-order SVD (HOSVD)\cite{de2000multilinear}.
The Tucker decomposition is an extension of SVD, and decomposes rating tensor $\mathcal{X}$ into a set of matrices and a small dense core tensor.
In the Tucker decomposition, $\bm{A}^{(1)}$, $\bm{A}^{(2)}$, and $\bm{A}^{(3)}$ contain the orthonormal vector called mode-1, mode-2 and mode-3 singular vectors, respectively.
Figure \ref{recommend:Tensor_decomposition} (b) illustrates the behavior of Tucker decomposition.
Third-order tensor $\mathcal{X}$ can also be approximated as Eqn. (\ref{recommend:eqn-tensor}), where $\mathcal{G}$ is also the core tensor, but unlike in CP decomposition, the core tensor is not a super diagonal tensor.
In Tucker decomposition, the rank value can be independently given for each mode of the rating tensor.
Herein we adopted \textsc{scikit-tensor}\cite{scikit-tensor} package to use tensor factorization methods.

\section{Datasets}
\label{section:datasets}
We employ three inorganic crystal structure databases of the ICSD, Powder Diffraction File (PDF) by the International Centre for Diffraction Data (ICDD)\cite{ICDD} and SpringerMaterials (SpMat)\cite{SpringerMaterials}. 
Only ICSD entries, including entries reported to show a partial occupancy behavior, are regarded as known CRCs.
Both ICDD and SpMat entries, which are not included in the ICSD, are used to validate the recommender systems.

\begin{table}[tbp]
\caption{
Elements included in the datasets of known CRCs.
Rare-earth elements shown in parentheses are excluded in the candidate quaternary and quinary compositions.
}
\label{recommend:elements1}
\begin{ruledtabular}
\begin{tabular}{ll}
Cations & Li,Na,K,Rb,Cs,Be,Mg,Ca,Sr,Ba,Zn,Cd,Hg, \\
        & B,Al,Sc,Y,La,Ga,In,Tl,Si,Ge,Sn,Pb,P,As,Sb,Bi \\
        & Ti,Zr,Hf,V,Nb,Ta,Cr,Mo,W,Mn,Tc,Re, \\
        & Fe,Ru,Os,Co,Rh,Ir,Ni,Pd,Pt,Cu,Ag,Au, \\
        & (Ce,Pr,Nd,Pm,Sm,Eu,Gd,Tb,Dy,Ho,Er,Tm,Yb,Lu) \\
Anions  & C,N,O,S,Se,Te,F,Cl,Br,I \\
\end{tabular}
\end{ruledtabular}
\end{table}

\begin{table}[tbp]
\caption{
Numbers of ICSD, ICDD and SpMat entries.
Numbers of candidate compositions are also shown.
ICSD entries are regarded as known CRCs.
In the rows of ICDD and SpMat databases, the numbers of entries that are not included in the ICSD are shown.
Numbers in the parentheses indicate the numbers of all entries.
Test dataset is a combined dataset of ICDD and SpMat entries.
Only in Sec. \ref{section:dft}, a combined dataset of ICSD, ICDD, and SpMat entries is used as the training data.
}
\label{recommend:n_compositions}
\begin{ruledtabular}
\begin{tabular}{cccc}
                                & Ternary & Quaternary & Quinary \\
\hline
ICSD                            & 9,313    & 7,742    & 1,321 \\
\multirow{2}{*}{ICDD}           & 2,369    & 2,647    & 639 \\
                                & (9,278)  & (7,864)  & (1,326) \\
\multirow{2}{*}{SpMat}          & 2,708    & 3,066    & 1,169 \\
                                & (10,461) & (8,141)  & (1,893) \\
\multirow{2}{*}{ICDD$+$SpMat}   & 4,134    & 4,961    & 1,616 \\
                                & (12,573) & (11,307) & (2,562) \\
\hline
Candidates   & 7,405,200 & 1,188,038,460 & 23,104,706,560
\end{tabular}
\end{ruledtabular}
\end{table}

To build a recommender system for ternary CRCs, only a set of ternary known CRCs is employed.
Candidate ternary compositions of A$_a$B$_b$X$_x$ are generated, where elements A and B correspond to a cation (66 possible) and element X corresponds to an anion (10 (10 possible). 
The possible cations and anions correspond to elements included in the dataset of known CRCs (Table \ref{recommend:elements1}).
One hundred seventy (170) integer sets $(a,b,x)$ that satisfy the condition of $\max (a,b,x) \leq 8$ and are included in the dataset of known CRCs are also considered.
Therefore, there are $66^2 \times 10 \times 170 = 7,405,200$ ternary compositions.
Of these, 9,313 CRCs are known.
The combined database of ICDD and SpMat has 4,134 ternary compositions that are not included in ICSD.
This corresponds to 0.056\% of the candidate compositions.
Table \ref{recommend:n_compositions} summarizes the numbers of database entries and candidate compositions.

In addition to ternary compositions, candidate quaternary A$_a$B$_b$C$_c$X$_x$ compositions are generated, where elements A, B and C denote a cation and element X denotes an anion. 
Integer sets $(a,b,c,x)$ satisfy $\max (a,b,c,x) \leq 20$.
The dataset of known quaternary CRCs includes 53 cations, 10 anions, and 798 integer sets.
Therefore the number of quaternary compositions is $53^3 \times 10 \times 798 = 1,188,038,460$, which includes 7,742 known CRCs.
The combined database of ICDD and SpMat includes 4,961 quaternary compositions ($4.2 \times 10^{-4}$\% of the candidate compositions) that are not the entries of ICSD.

Candidate quinary A$_a$B$_b$C$_c$D$_d$X$_x$ compositions are also generated, where elements A, B, C and D denote cations and element X denotes an anion. 
Integer sets $(a,b,c,d,x)$ satisfy $\max (a,b,c,d,x) \leq 20$.
The dataset of known quinary CRCs includes 52 cations, 10 anions, 316 integer sets.
Therefore, the number of quaternary compositions is $52^4 \times 10 \times 316 = 23,104,706,560$, where only 1,321 known CRCs are included.
The combined database of ICDD and SpMat contains 1,616 quinary compositions ($7.0 \times 10^{-6}$\% of the candidate compositions) that are not ICSD entries.

\section{Matrix-based recommender system}
\subsection{Rating matrix representations}
The performance of a recommender system strongly depends on the representation of the composition dataset as a rating matrix or tensor.
Experts' knowledge is required to introduce a good representation.
In particular, a composition dataset must be transformed into only two sets of features, which corresponds to users and items in a user-item rating matrix.
The ratings of missing elements are approximately predicted on the basis of feature similarity given by the representation.

\begin{figure}[tbp]
\begin{center}
\includegraphics[width=0.9\linewidth,clip]{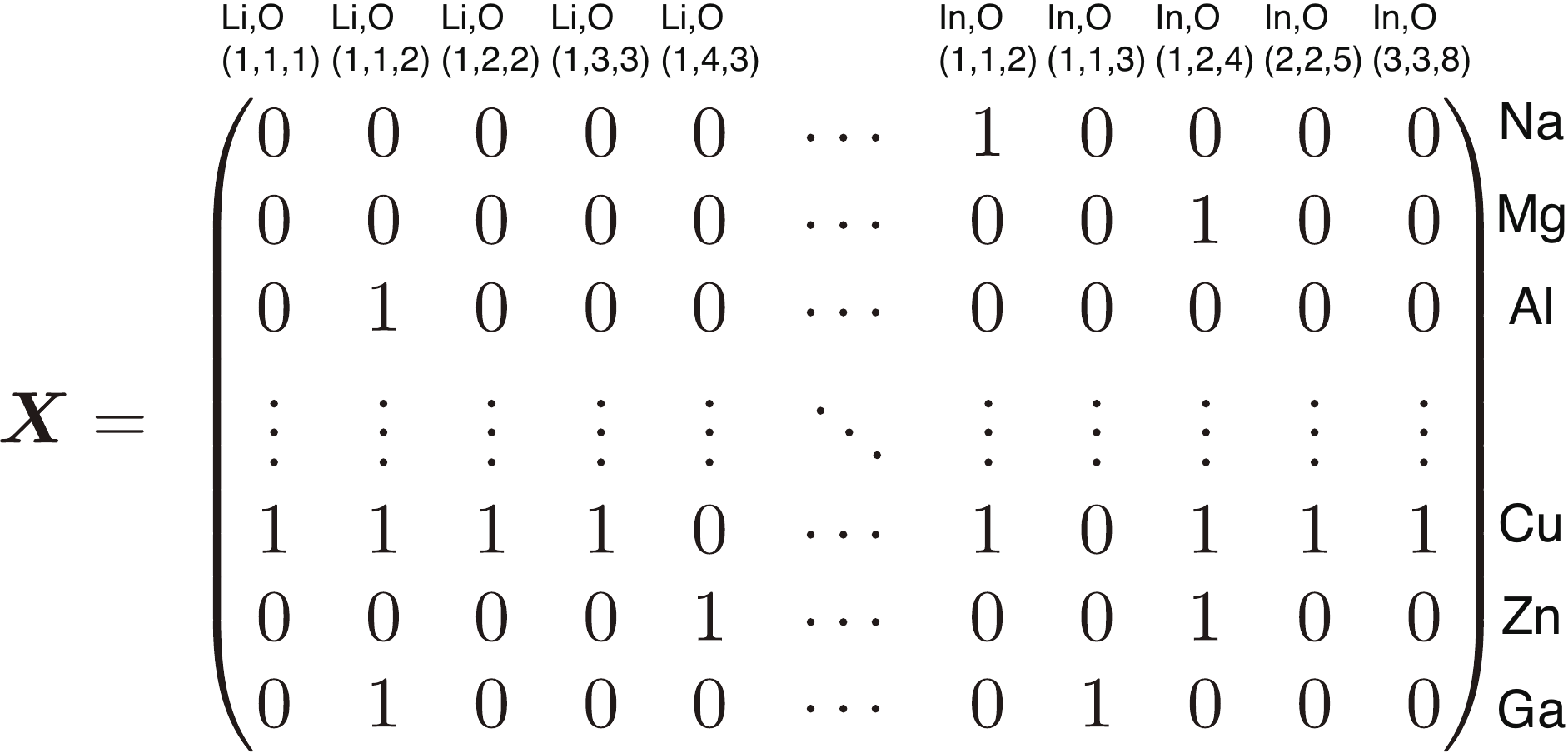} 
\caption{
Schematic illustration of a matrix representation for a composition dataset.
}
\label{recommend:MatrixRep}
\end{center}
\end{figure}

We introduce three kinds of matrix representations for the ternary composition dataset.
A composition is decomposed into two feature sets in the following three ways.
\begin{enumerate}
\item \{A\} and \{B, X, $(a,b,x)$\}
\item \{A, X\} and \{B, $(a,b,x)$\}
\item \{A, B\} and \{X, $(a,b,x)$\}
\end{enumerate}
Each of feature sets corresponds to the row or column of the rating matrix.
For example, the first matrix representation is schematically illustrated in Fig. \ref{recommend:MatrixRep}.
In the first representation, the row corresponds to cation A.
The column corresponds to the combination of cation B, anion X, and integer set $(a,b,x)$.
Each composition is expressed by the collection of features included in the row and column.
Only the values of the rating matrix elements corresponding to known CRCs are set to unity.
The other compositions are classified as either currently unknown CRCs or non-existent compositions, and their values are set to zero in the original rating matrix.

\subsection{Discovery performance of unknown CRCs}
The performance of matrix factorization based the recommender systems to discover currently unknown ternary CRCs are evaluated.
Only 4,134 of 7,405,200 compositions (0.056\%) are used as the test dataset of known CRCs (test CRCs). 
The test CRCs correspond to ICDD and SpMat entries that are not included in ICSD.
The low percentage obviously confirms that discovering ICDD and SpMat entries by random sampling is not efficient.

\begin{figure}[tbp]
\begin{center}
\includegraphics[width=\linewidth,clip]{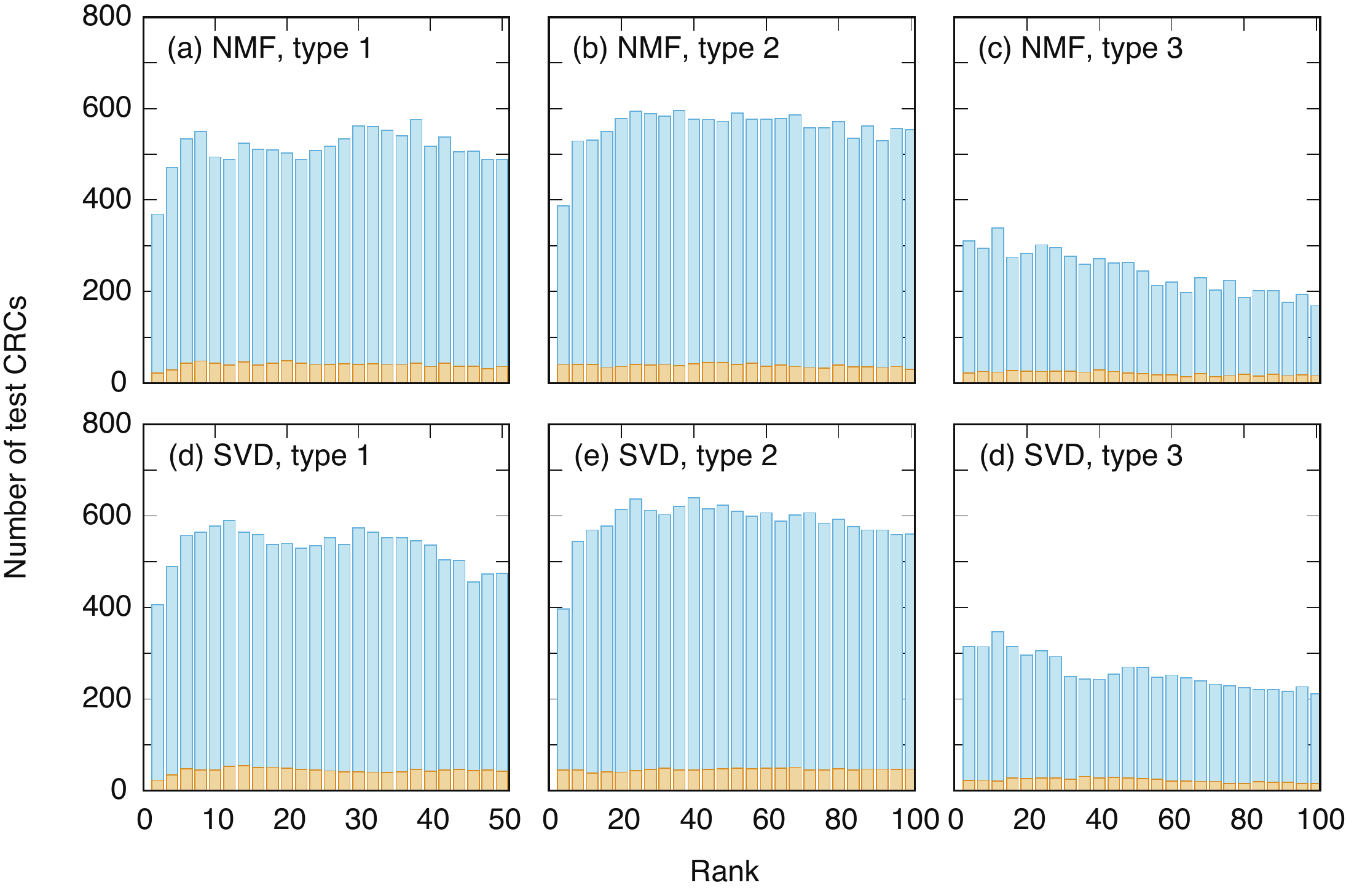} 
\caption{
Numbers of test CRCs included in the top 100 and 3000 compositions of NMF recommender systems for (a) type-1, (b) type-2 and (c) type-3 matrix representations.
Numbers of test CRCs included in the top 100 and 3000 compositions of SVD recommender systems for (d) type-1, (e) type-2 and (f) type-3 matrix representations.
Horizontal axis indicates a given rank for NMF and SVD.
Orange and blue bars denote the top 100 and 300 compositions, respectively.
}
\label{recommend:ResultMF}
\end{center}
\end{figure}

Figure \ref{recommend:ResultMF} shows the rank dependence of the number of test CRCs found in the top 3000 compositions of NMF and SVD recommender systems.
The rank dependences of NMF and SVD are qualitatively similar but weak.
They also exhibit similar matrix representation dependences.
The performance of discovering test CRCs of type-1 and type-2 matrix representations is much better than that of type-3 matrix representations.
This implies that it is important to consider the cation and anion combination as a feature when building a rule to distinguish CRCs and non-existent compositions.

When adopting type-1 and type-2 matrix representations, 
NMF recommender systems with type-1 and type-2 matrix representations show the best performance at $r = 38$ and $r = 24$, respectively, indicating that the maximum number of test CRCs are included in the top 3000 compositions of the recommender system.
For type-1 and type-2 representations, 576 and 594 test CRCs (19.2 and 19.8\%) are found in the top 3000 compositions, respectively.
These percentages are much larger than the number of test CRCs discovered by random sampling (0.056\%).
A much larger discovery rate of test CRCs is seen in the sampling of top 100 compositions. 
The numbers of test CRCs are 44 and 41 (44.0 and 41.0\%), respectively.
SVD recommender systems with type-1 and type-2 matrix representations show the best performance at $r=12$ and $r=40$, respectively. 
The top 3000 compositions include 590 and 640 test CRCs (19.7 and 21.3\%), respectively.
Additionally, 53 and 45 of the top 100 compositions (53.0 and 45.0\%) correspond to test CRCs.
These results indicate that SVD recommender systems perform slightly better than NMF recommender systems.

\section{Tensor decomposition-based recommender system}
\subsection{Rating tensor representation}

\begin{figure}[tbp]
\begin{center}
\includegraphics[width=0.7\linewidth,clip]{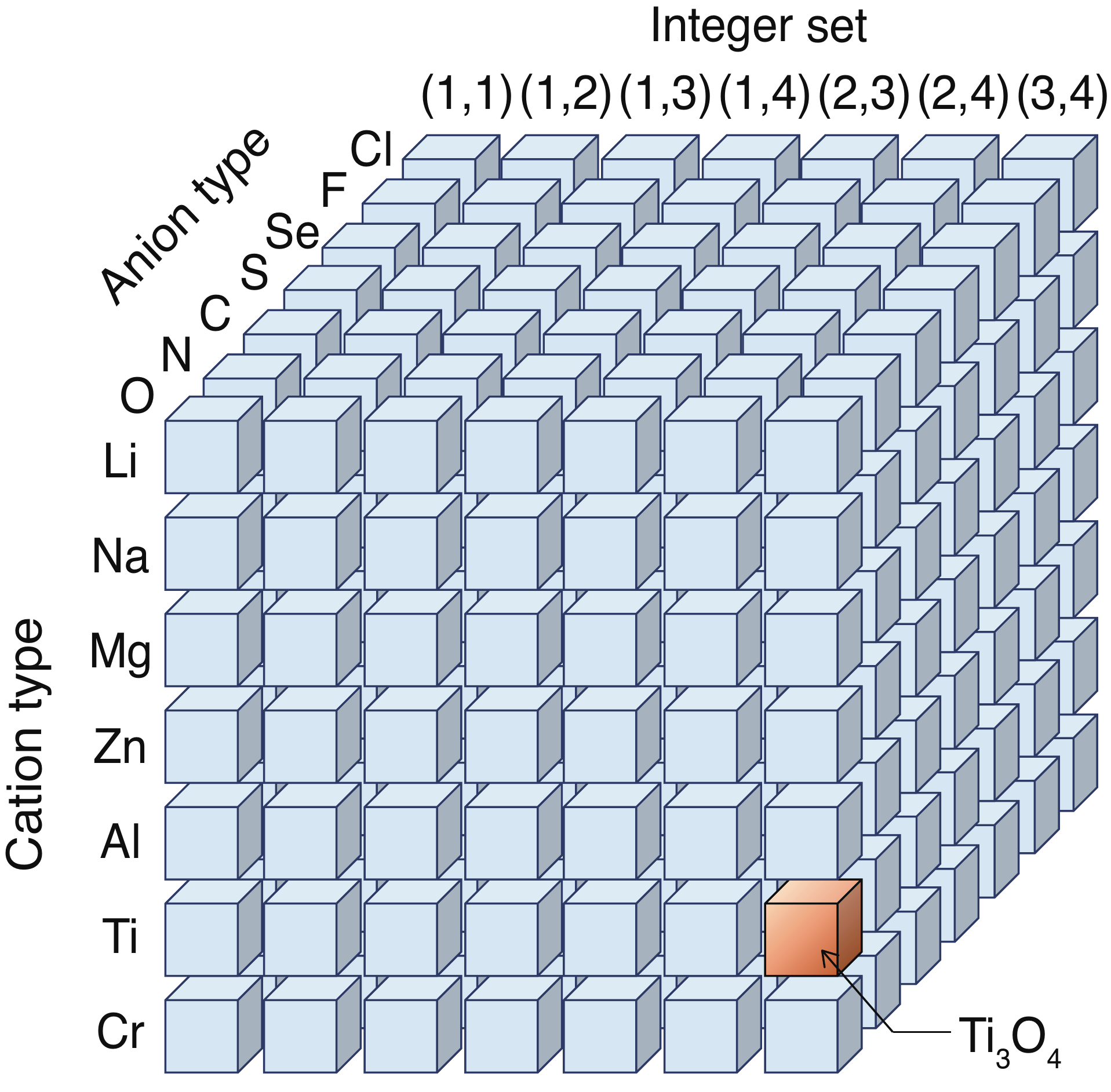} 
\caption{
Schematic illustration of the tensor representation of a composition dataset.
As a simple example, a tensor representation for binary compositions is illustrated.
In a tensor element highlighted by the orange block, modes for cation type, anion type and integer set indicate Ti, O and $(3,4)$, respectively. 
Hence, this tensor element represents the composition of Ti$_3$O$_4$.
}
\label{recommend:TensorRep}
\end{center}
\end{figure}

Figure \ref{recommend:TensorRep} illustrates the tensor representation used in this study.
A mode of the rating tensor means each cation type A, cation type B, anion type X and integer set $\{a, b, x\}$ for ternary compositions A$_a$B$_b$X$_x$.
Therefore, the rating tensor for ternary compositions is a (66, 66, 10, 198)-dimensional tensor.
Similar to the matrix representations, the composition is expressed by the collections of features included in all modes, corresponding to an element of the rating tensor.
Only the values of the rating tensor elements corresponding to known CRCs are set to unity. 
The values of tensor elements corresponding to the other compositions are assumed to be zero in the original rating tensor.
By representing quaternary and quinary compositions in tensor forms as well as ternary compositions, the numbers of modes of the rating tensor correspond to five and six for quaternary and quinary compositions, respectively.

\subsection{Discovery performance of unknown CRCs}

\begin{figure}[tbp]
\begin{center}
\includegraphics[width=\linewidth,clip]{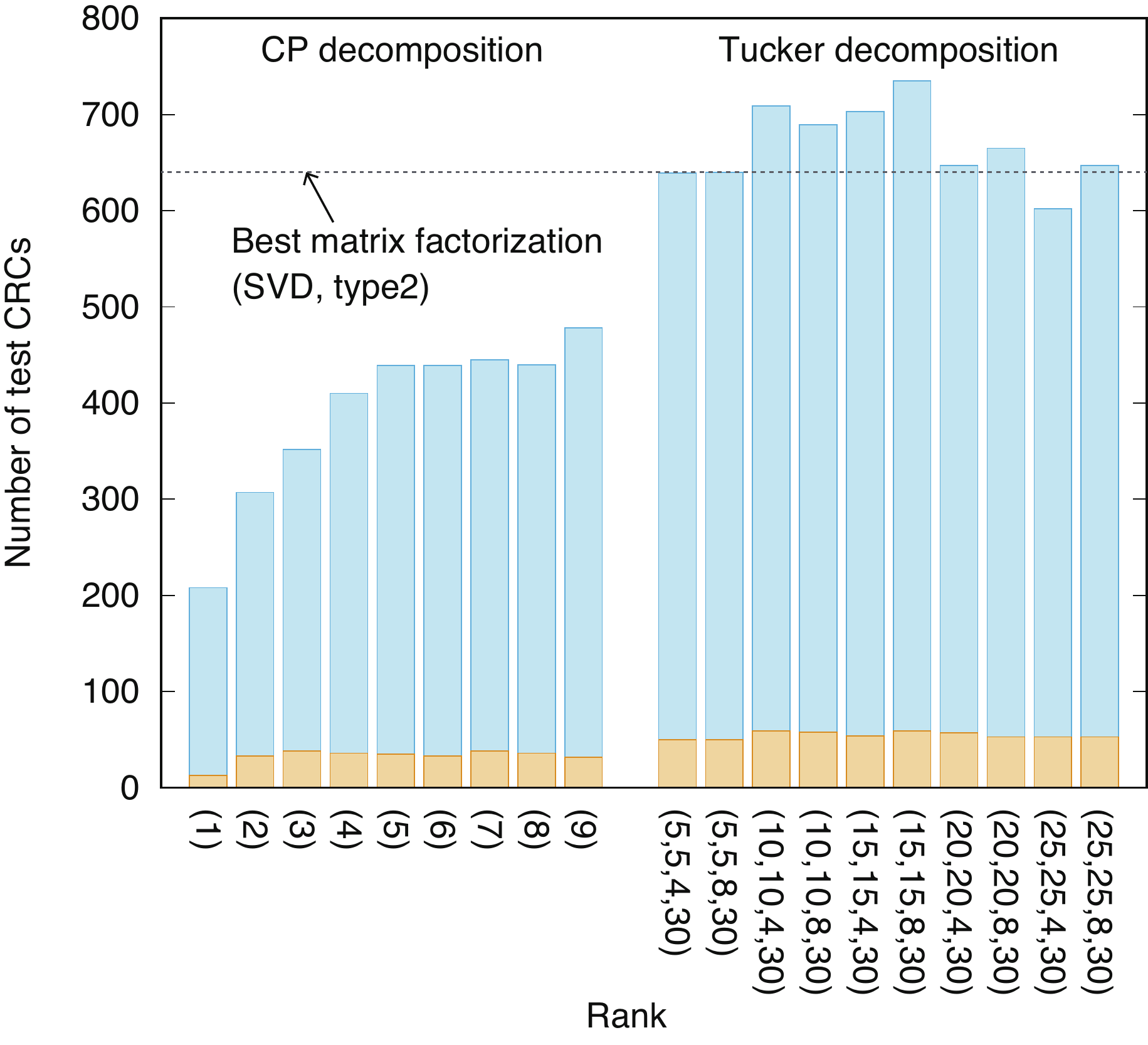} 
\caption{
Rank dependence of the number of test CRCs included in the top 100 and 3000 compositions of the CP decomposition recommender systems.
Rank dependence of the number of test CRCs included in the top 100 and 3000 compositions of the Tucker decomposition recommender systems.
Here only the performance of Tucker decomposition with a rank value of 30 for the composition integer set is shown.
Orange and blue bars show the top 100 and 3000 compositions, respectively.
}
\label{recommend:ResultTensor}
\end{center}
\end{figure}

Figure \ref{recommend:ResultTensor} shows the number of test CRCs included in the top 3000 ternary compositions of the ranking by CP decomposition.
The horizontal axis indicates a given rank for CP decomposition.
Since only a single value of rank is given in CP decomposition, the maximum value of rank is identical to the number of anion types included in the ternary composition dataset.
The CP recommender system with $r=9$ shows the best performance.
That is, 478 of the top 3000 compositions (15.9\%) correspond to test CRCs.
In the top 100 compositions, 32 test CRCs (32.0\%) are found.
However, CP decompositions detect fewer test CRCs than the best matrix decomposition-based recommender system. 
This is attributed to the fact that only a single value of rank is given in CP decomposition.

Figure \ref{recommend:ResultTensor} shows the number of test CRCs included in the top 3000 compositions of the ranking by Tucker decomposition.
The candidate values of ranks for the cation mode, anion mode, and integer set mode are given by arithmetic sequences of \{5, 10, 15, 20, 25\}, \{4, 8\} and \{10, 20, 30, $\dots$, 100\}, respectively.
The optimal set of ranks is obtained by a grid search using these candidates. 
This means that the performance of discovering test CRCs is examined for 100 combinations of rank values.
The Tucker decomposition recommender system with ranks of (15, 15, 8, 30) shows the best performance; 735 test CRCs (24.5\%) are found in the top 3000 compositions.
In the top 100 compositions, 59 CRCs (59.0\%) are found in the test dataset.
The Tucker recommender system detects more test CRCs than the best matrix-based recommender system. 

\begin{figure}[htbp]
\begin{center}
\includegraphics[width=0.8\linewidth,clip]{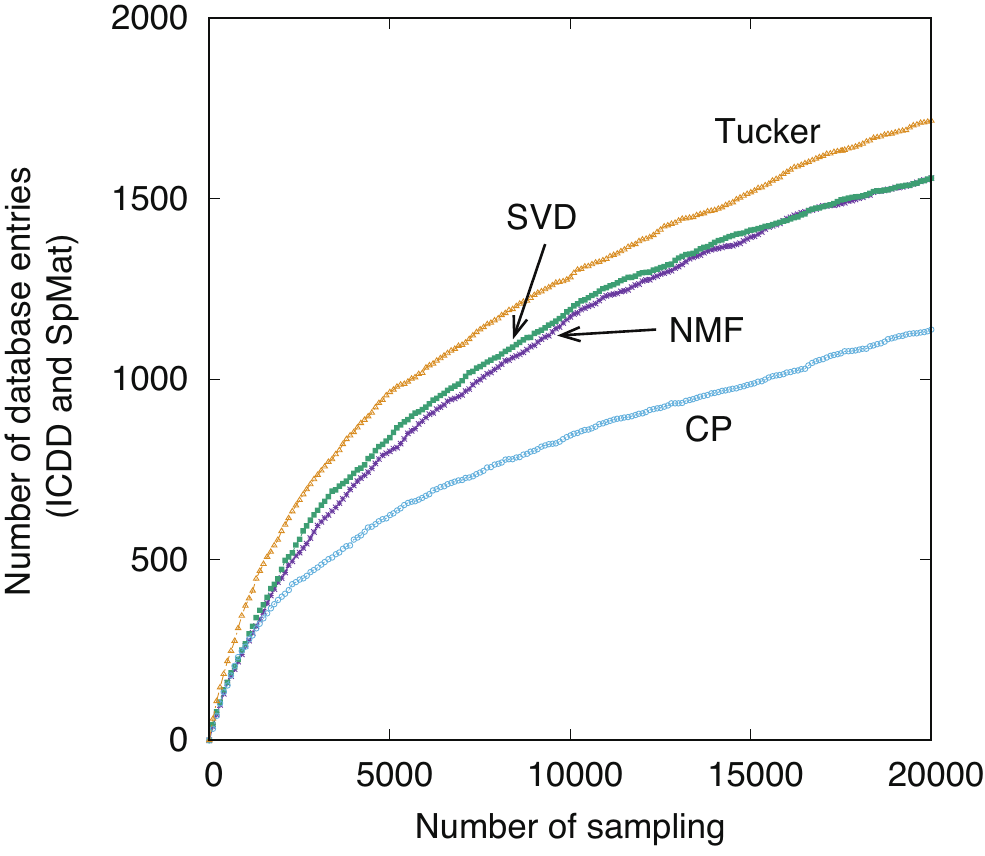} 
\caption{
Comparison of the dependence of the number of test CRCs as a function of the number of samples obtained from the rankings by the best NMF, best SVD, best CP decomposition, and best Tucker decomposition recommender systems.
Derivatives of these curves correspond to the discovery rate of the test CRCs.
}
\label{recommend:ResultBestRecommend}
\end{center}
\end{figure}

Figure \ref{recommend:ResultBestRecommend} compares the number of test CRCs found by the best NMF, best SVD, best CP decomposition and best Tucker decomposition recommender systems.
The optimal Tucker decomposition recommender system exhibits the best performance among the four systems.
This result is due to the features of Tucker decomposition.
Compared with matrix factorizations, Tucker decompositions naturally consider the low-rank data-structure of the rating tensor related to two or more modes, although only correlations between the two modes of a given representation are in matrix factorizations. 
Multiple rank values can be given in a Tucker decomposition, whereas a CP decomposition has a single value for rank.
Therefore, Tucker decompositions allow a hidden low-rank data-structure included in the composition dataset to be extracted more flexibly than matrix factorizations and CP decompositions.

The Tucker decomposition should be the best approach to discover currently unknown CRCs.
Therefore, the performance to discover currently unknown quaternary and quinary CRCs using the Tucker decomposition is examined.
The performances for quaternary and quinary systems are particularly important features required for recommender systems because the numbers of candidate compositions and currently unknown CRCs should be exponentially huge.
As described in Sec. \ref{section:datasets}, only 7,742 and 4,961 of $ 1.2 \times 10^9$ (6.5 $\times 10^{-4}$ and 4.2 $\times 10^{-4}$\%) quaternary compositions are ICSD and ICDD+SpMat entries, respectively.
Even for quinary compositions, only 1,321 and 1,616 of $ 2.3 \times 10^{10}$ (5.7 $\times 10^{-6}$ and 7.0 $\times 10^{-6}$\%) compositions correspond to ICSD and ICDD+SpMat entries, respectively.

\begin{figure}[tbp]
\begin{center}
\includegraphics[width=\linewidth,clip]{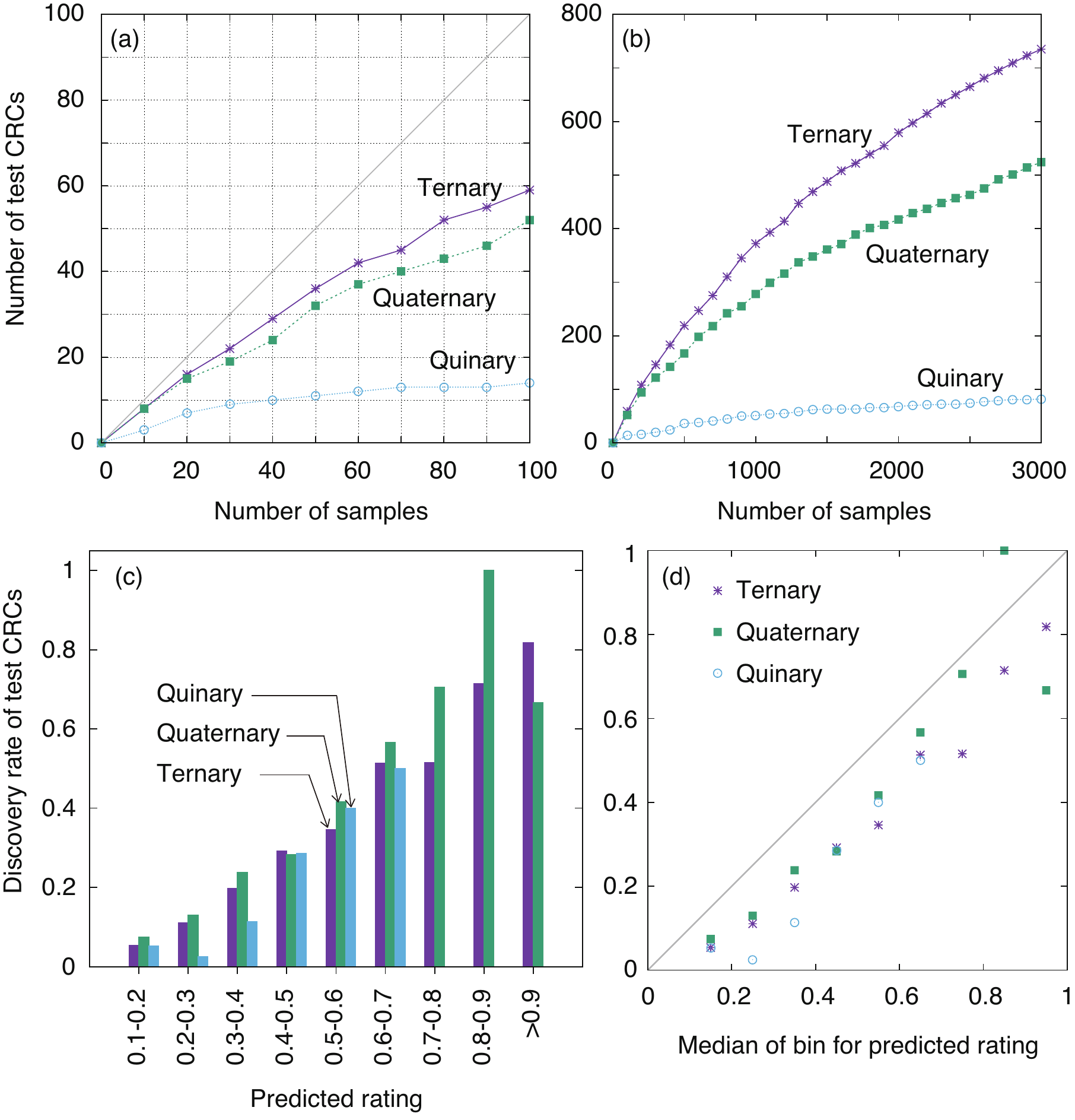} 
\caption{
Dependence of the number of test quaternary and quinary CRCs on the number of samples by the Tucker decomposition recommender system up to (a) 100 and (b) 3000 samples.
Diagonal line in (a) indicates the performance of the ideal sampling (i.e. the number of test CRCs is identical to the number of samples).
(c) Dependence of the discovery rate of test CRCs on the predicted rating for ternary, quaternary and quinary compositions.
(d) Relationship between the predicted rating and the discovery rate of the test CRCs.
Horizontal axis indicates the median of bin for the predicted rating.
}
\label{recommend:ResultTucker2}
\end{center}
\end{figure}

Figures \ref{recommend:ResultTucker2} (a) and (b) show the performance to determine test ternary, quaternary and quinary CRCs of the Tucker recommender system.
The discovery rate gradually decreases as the number of samples all for ternary, quaternary and quinary compositions increases.
In the top 20 compositions, 16 and 15 compositions (80 and 75\%) correspond to test CRCs for ternary and quaternary compositions, respectively.
The discovery rate of quaternary test CRCs is higher than expected.
The rate is close to that of ternary test CRCs.
For quaternary compositions, 52 and 524 CRCs are found in the top 100 and 3000 compositions (52.0 and 17.5\%), respectively.
The discovery rates for the top 20, 100, and 3000 compositions are approximately 180,000, 120,000, and 40,000 times larger than the random sampling discovery rate, respectively.
Even for quinary compositions, seven test CRCs (35\%) are found in the top 20 compositions.
In addition, 14 and 82 CRCs are found in the top 100 and 3000 compositions (14.0 and 2.7\%), respectively.
Although the discovery rate of quinary test CRCs is lower than those of ternary and quaternary test CRCs, it is much larger than the random sampling.
In particular, the discovery rate for the top 100 compositions is astonishingly large despite the fact that few known quinary CRCs are included in the quinary composition dataset.

Figures \ref{recommend:ResultTucker2} (c) and (d) show the relationship between the predicted rating and the discovery rate of test CRCs.
Figure \ref{recommend:ResultTucker2} (c) shows the histogram of the discovery rate of the test CRCs obtained with a bin width of predicted rating of 0.1.
Figure \ref{recommend:ResultTucker2} (d) replots the relationship between the median of the bin and discovery rate from the histogram of the discovery rate.
For ternary, quaternary and quinary compositions, the discovery rate of test CRCs is almost proportional to the median of the bin for predicted rating, and slightly smaller than the predicted rating.
This is strong evidence that the predicted rating can be regarded as a figure of merit to identify currently unknown CRCs.
This may also indicate that the predicted rating is almost identical to the discovery rate of currently unknown CRCs because test CRCs can be regarded as a part of currently unknown CRCs.
In addition, quinary compositions with a predicted rating exceeding 0.7 are not observed.
Consequently, the discovery rate of test CRCs is smaller than that of the ternary and quaternary test CRCs.
This is mainly ascribed to the lack of quinary known CRCs.
The number of quinary known CRCs is inadequate to predict currently unknown quinary CRCs.

\begin{figure*}[tbp]
\begin{center}
\includegraphics[width=\linewidth,clip]{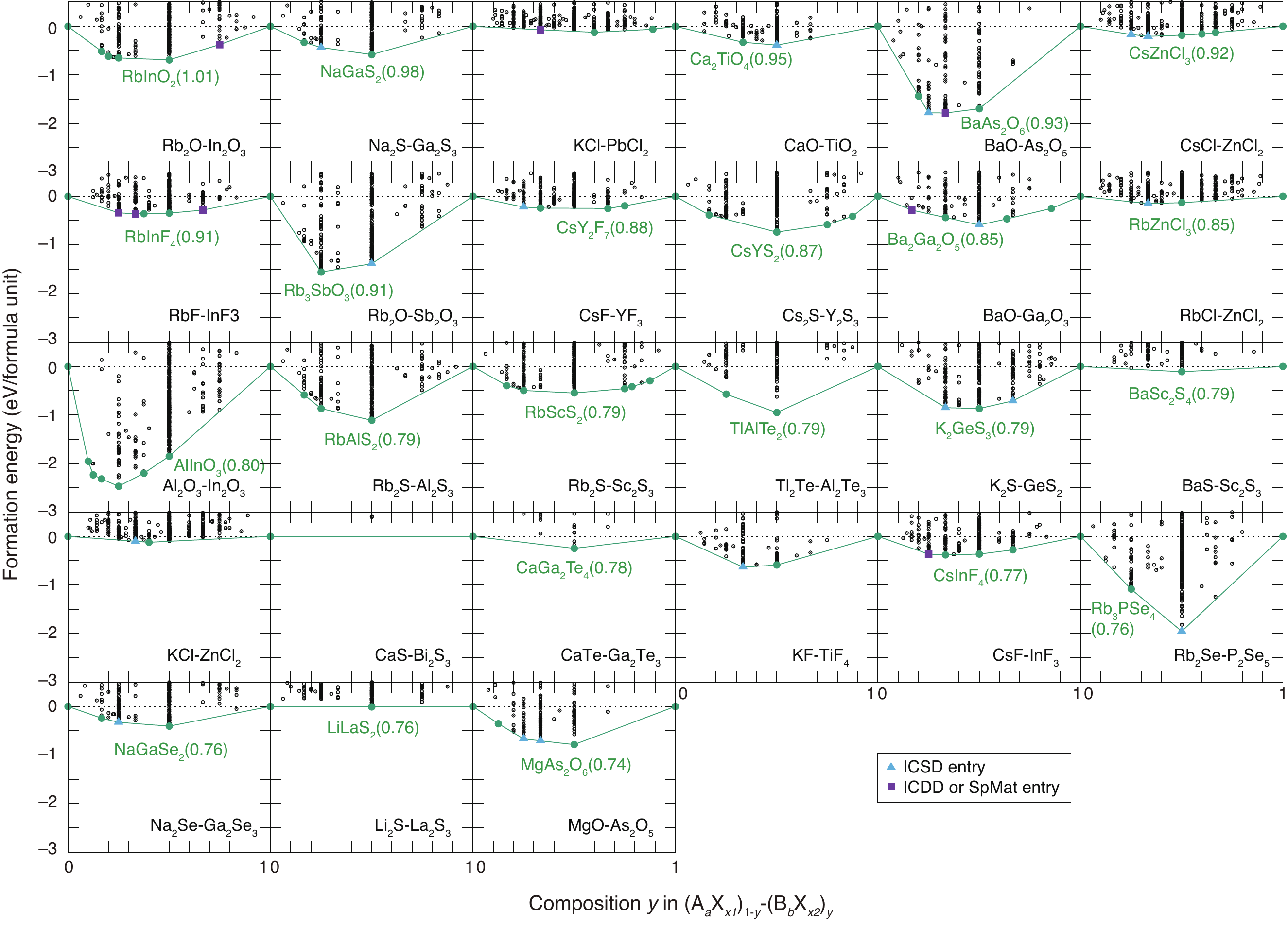} 
\caption{
Stable compounds on the convex hull of the formation energy computed by the DFT calculations for the 27 pseudo-binary systems containing the top 27 compositions with high predicted ratings.
Compositions with high predicted ratings are shown along with their predicted ratings as the values in parentheses.
Closed triangles and squares denote the stable compounds corresponding ICSD entries and ICDD+SpMat entries, respectively.
}
\label{recommend:dft1}
\end{center}
\end{figure*}

\begingroup
\squeezetable
\begin{table}[htbp]
\caption{
Compositions with a predicted rating of more than 0.2 included in pseudo-binary systems for the top 27 compositions with a high predicted rating.
Stability of the compositions by the DFT calculation is also shown.
}
\label{recommend:dft-table}
\begin{ruledtabular}
\begin{tabular}{ccc}
Composition & Predicted rating & Stability \\
\hline
RbInO$_2$       & 1.01 & $\bigcirc$ \\
Rb$_3$InO$_3$   & 0.64 & $\bigcirc$ \\
RbIn$_5$O$_8$   & 0.20 & $-$ \\
\hline
NaGaS$_2$       & 0.98 & $\bigcirc$ \\
NaGa$_5$S$_8$   & 0.21 & $-$ \\
\hline
KPbCl$_3$       & 0.97 & $-$ \\
\hline
Ca$_2$TiO$_4$   & 0.95 & $\bigcirc$ \\
CaTi$_3$O$_7$   & 0.29 & $-$ \\
Ca$_3$TiO$_5$   & 0.21 & $-$ \\
\hline
BaAs$_2$O$_6$   & 0.93 & $\bigcirc$ \\
\hline
CsZnCl$_3$      & 0.92 & $\bigcirc$ \\
CsZn$_2$Cl$_5$  & 0.25 & $\bigcirc$ \\
\hline
RbInF$_4$       & 0.91 & $\bigcirc$ \\
\hline
Rb$_3$SbO$_3$   & 0.91 & $\bigcirc$ \\
\hline
CsY$_2$F$_7$    & 0.87 & $\bigcirc$ \\
Cs$_2$YF$_5$    & 0.44 & $\bigcirc$ \\
CsYF$_4$        & 0.38 & $-$ \\
\hline
CsYS$_2$        & 0.87 & $\bigcirc$ \\
\hline
Ba$_2$Ga$_2$O$_5$ & 0.85 & $\bigcirc$ \\
BaGa$_4$O$_7$   & 0.55 & $-$ \\
\hline
RbZnCl$_3$      & 0.85 & $\bigcirc$ \\
RbZn$_2$Cl$_5$  & 0.23 & $-$ \\
\hline
AlInO$_3$       & 0.80 & $\bigcirc$ \\
\hline
RbAlS$_2$       & 0.79 & $\bigcirc$ \\
Rb$_3$AlS$_3$   & 0.43 & $\bigcirc$ \\
RbAl$_5$S$_8$   & 0.26 & $-$ \\
\hline
RbScS$_2$       & 0.79 & $\bigcirc$ \\
\hline
TlAlTe$_2$      & 0.79 & $\bigcirc$ \\
TlAl$_5$Te$_8$  & 0.23 & $-$ \\
\hline
K$_2$GeS$_3$    & 0.79 & $\bigcirc$ \\
K$_6$Ge$_2$S$_7$ & 0.25 & $-$ \\
\hline
BaSc$_2$S$_4$   & 0.79 & $\bigcirc$ \\
\hline
KZnCl$_3$       & 0.79 & $-$ \\
KZn$_2$Cl$_5$   & 0.21 & $-$ \\
\hline
CaBi$_2$S$_4$   & 0.78 & $-$ \\
Ca$_2$Bi$_2$S$_5$ & 0.28 & $-$ \\
\hline 
CaGa$_2$Te$_4$  & 0.78 & $\bigcirc$ \\
\hline
K$_3$TiF$_7$    & 0.78 & $-$ \\
KTiF$_5$        & 0.26 & $\bigcirc$ \\
\hline
CsInF$_4$       & 0.77 & $\bigcirc$ \\
CsIn$_2$F$_7$   & 0.59 & $\bigcirc$ \\
Cs$_2$InF$_5$   & 0.49 & $\bigcirc$ \\
\hline
Rb$_3$PSe$_4$   & 0.76 & $\bigcirc$ \\
\hline
NaGaSe$_2$      & 0.76 & $\bigcirc$ \\
\hline
LiLaS$_2$       & 0.76 & $\bigcirc$ \\
LiLa$_5$S$_8$   & 0.29 & $-$ \\
\hline
MgAs$_2$O$_6$   & 0.74 & $\bigcirc$ \\
\end{tabular}
\end{ruledtabular}
\end{table}
\endgroup

\section{Discovery of new compounds}
\label{section:dft}
Finally, the discovery of new ternary compounds is demonstrated using a combination of the Tucker recommender system and DFT calculations.
A Tucker decomposition recommender system is constructed using all of the available entries of the ternary compositions included in ICSD, ICDD and SpMat as known CRCs.
The rank values are the same values as those optimized above.
The existence of stable compounds at compositions with high predicted ratings is examined by evaluating the phase stability of the corresponding pseudo-binary systems using DFT calculations.
DFT calculations were performed only for pseudo-binary systems containing the top 27 compositions composed only of elements in Groups 1, 2, 3, 4, 5, 6, 12, 13, 14, 15, 16 and 17 in the periodic table.
For each pseudo-binary system, DFT calculations were performed for all possible prototype structures included in the ICSD.
Totally, the number of DFT calculations was 13,274.

All DFT calculations were performed using the plane-wave basis projector augmented wave (PAW) method\cite{PAW1,PAW2} within the Perdew--Burke--Ernzerhof exchange-correlation functional\cite{GGA:PBE96} as implemented in the \textsc{vasp} code\cite{VASP1,VASP2}.
The cutoff energy was set to 400 eV.
The total energy converged to less than 10$^{-3}$ meV.
The atomic positions and lattice constants were optimized until the residual forces became less than $10^{-2}$ eV/\AA.

Figure \ref{recommend:dft1} shows the stable compounds or CRCs obtained by the DFT calculations for the 27 pseudo-binary systems.
The predicted ratings of the top 27 compositions range from 0.74 to 1.01.
Among the 27 compositions, most (23 compositions) are located on the convex hull, meaning that they are CRCs according to the DFT calculations. 
In other words, 85\% of the recommended compositions can be regarded as new CRCs, demonstrating that the predicted rating can be regarded as the discovery rate of new CRCs as described above.
Table \ref{recommend:dft-table} shows the compositions with predicted ratings greater than $0.2$ included in the 27 pseudo-binary systems and their stability obtained by the DFT calculations.
In addition to the top 27 compositions, some other compositions can be CRCs depending on the predicted rating.
In compositions with a predicted rating of 0.2--0.4, only two of the 14 compositions (14\%) correspond to CRCs.
On the other hand, four of the five compositions (80\%), 13 of the 16 compositions (81\%) and 11 of the 12 compositions (92\%) are CRCs with predicted ratings of 0.4--0.6, 0.6--0.8 and 0.8--1.0, respectively.
In other words, compounds with a predicted rating exceeding 0.4 have a high probability of being a CRC.

The remaining 15\% of recommended compositions have a high predicted rating, but do not correspond to the CRCs in the DFT calculations.
Plausible explanations include the following.
(1) A stable or metastable compound does not truly exist at the composition.
(2) A metastable compound can be found at the composition.
(3) A stable compound with a structure other than the prototype structures is observed.

In addition, stable compounds can be found at the other compositions with a high predicted rating.
The compositions are classified into two types.
The first corresponds to compositions omitted in the candidate dataset, which do not satisfy the condition of $\max(a, b, x) \leq 8$.
This type may be predicted with a high rating if a larger maximum value of the integer is considered.
The other corresponds to compositions with a low predicted rating.
They may be the true DFT stable compound when a low rating is predicted due to the lack of known CRCs similar to the corresponding composition. 
On the other hand, they may be artificial stable compounds attributed to the use of only the prototype structures for the DFT calculations.

\section{Conclusion}
Herein we apply widely-used recommender system approaches based on NMF, SVD, CP decomposition, and Tucker decomposition to predict currently unknown CRCs.
The present study indicates that the performance of the recommender systems depends on not only the given rank corresponding to a low-rank structure hidden in the rating matrix and tensor but also on the representation of rating matrix and tensor.
Among the considered approaches, a Tucker decomposition recommender system shows the best discovery rate of CRCs.
For ternary and quaternary compositions, at least approximately 60 and 50 of the top 100 recommended compositions are CRCs, respectively.

We also demonstrate the discovery of new ternary compounds using a combination of a Tucker recommender system and DFT calculations.
Among the top 27 compositions whose predicted ratings ranging 0.74 to 1.01, 23 compositions correspond to CRCs. 
This suggests that the discovery rate of currently unknown CRCs is surprisingly high (85\%).
Moreover, the predicted ratings reveal additional compositions that may also be CRCs.
The discovery rate of CRCs using DFT calculations is almost identical to the predicted rating of the recommender system. 
Consequently, the predicted rating can be regarded as a quantitative figure of merit to determine currently unknown CRCs.
Employing this figure of merit may significantly accelerate the discovery of currently unknown CRCs.

\begin{acknowledgments}
This work was supported by PRESTO, JST, and a Grant-in-Aid for Scientific Research on Innovative Areas ``Nano Informatics" (Grant No. 25106005) from the Japan Society for the Promotion of Science (JSPS).
AS, HH and IT were supported by the ``Materials Research by Information Integration" Initiative (MI$^2$I) from the Japan Science and Technology Agency.
IT was supported by a Grant-in-Aid for Scientific Research (A) from JSPS.
\end{acknowledgments}

\bibliography{recommend}

\begin{thebibliography}{23}%
\makeatletter
\providecommand \@ifxundefined [1]{%
 \@ifx{#1\undefined}
}%
\providecommand \@ifnum [1]{%
 \ifnum #1\expandafter \@firstoftwo
 \else \expandafter \@secondoftwo
 \fi
}%
\providecommand \@ifx [1]{%
 \ifx #1\expandafter \@firstoftwo
 \else \expandafter \@secondoftwo
 \fi
}%
\providecommand \natexlab [1]{#1}%
\providecommand \enquote  [1]{``#1''}%
\providecommand \bibnamefont  [1]{#1}%
\providecommand \bibfnamefont [1]{#1}%
\providecommand \citenamefont [1]{#1}%
\providecommand \href@noop [0]{\@secondoftwo}%
\providecommand \href [0]{\begingroup \@sanitize@url \@href}%
\providecommand \@href[1]{\@@startlink{#1}\@@href}%
\providecommand \@@href[1]{\endgroup#1\@@endlink}%
\providecommand \@sanitize@url [0]{\catcode `\\12\catcode `\$12\catcode
  `\&12\catcode `\#12\catcode `\^12\catcode `\_12\catcode `\%12\relax}%
\providecommand \@@startlink[1]{}%
\providecommand \@@endlink[0]{}%
\providecommand \url  [0]{\begingroup\@sanitize@url \@url }%
\providecommand \@url [1]{\endgroup\@href {#1}{\urlprefix }}%
\providecommand \urlprefix  [0]{URL }%
\providecommand \Eprint [0]{\href }%
\providecommand \doibase [0]{http://dx.doi.org/}%
\providecommand \selectlanguage [0]{\@gobble}%
\providecommand \bibinfo  [0]{\@secondoftwo}%
\providecommand \bibfield  [0]{\@secondoftwo}%
\providecommand \translation [1]{[#1]}%
\providecommand \BibitemOpen [0]{}%
\providecommand \bibitemStop [0]{}%
\providecommand \bibitemNoStop [0]{.\EOS\space}%
\providecommand \EOS [0]{\spacefactor3000\relax}%
\providecommand \BibitemShut  [1]{\csname bibitem#1\endcsname}%
\let\auto@bib@innerbib\@empty
\bibitem [{\citenamefont {Deml}\ \emph {et~al.}(2016)\citenamefont {Deml},
  \citenamefont {O'Hayre}, \citenamefont {Wolverton},\ and\ \citenamefont
  {Stevanovi\ifmmode~\acute{c}\else \'{c}\fi{}}}]{PhysRevB.93.085142}%
  \BibitemOpen
  \bibfield  {author} {\bibinfo {author} {\bibfnamefont {A.~M.}\ \bibnamefont
  {Deml}}, \bibinfo {author} {\bibfnamefont {R.}~\bibnamefont {O'Hayre}},
  \bibinfo {author} {\bibfnamefont {C.}~\bibnamefont {Wolverton}}, \ and\
  \bibinfo {author} {\bibfnamefont {V.}~\bibnamefont
  {Stevanovi\ifmmode~\acute{c}\else \'{c}\fi{}}},\ }\href {\doibase
  10.1103/PhysRevB.93.085142} {\bibfield  {journal} {\bibinfo  {journal} {Phys.
  Rev. B}\ }\textbf {\bibinfo {volume} {93}},\ \bibinfo {pages} {085142}
  (\bibinfo {year} {2016})}\BibitemShut {NoStop}%
\bibitem [{\citenamefont {Bergerhoff}\ and\ \citenamefont
  {Brown}(1987)}]{bergerhoff1987crystal}%
  \BibitemOpen
  \bibfield  {author} {\bibinfo {author} {\bibfnamefont {G.}~\bibnamefont
  {Bergerhoff}}\ and\ \bibinfo {author} {\bibfnamefont {I.~D.}\ \bibnamefont
  {Brown}},\ }in\ \href@noop {} {\emph {\bibinfo {booktitle} {Crystallographic
  Databases}}},\ \bibinfo {editor} {edited by\ \bibinfo {editor} {\bibfnamefont
  {F.~H.}\ \bibnamefont {Allen~et al.}}}\ (\bibinfo  {publisher} {International
  Union of Crystallography, Chester},\ \bibinfo {year} {1987})\BibitemShut
  {NoStop}%
\bibitem [{\citenamefont {Hautier}\ \emph
  {et~al.}(2010{\natexlab{a}})\citenamefont {Hautier}, \citenamefont {Fischer},
  \citenamefont {Jain}, \citenamefont {Mueller},\ and\ \citenamefont
  {Ceder}}]{hautier2010finding}%
  \BibitemOpen
  \bibfield  {author} {\bibinfo {author} {\bibfnamefont {G.}~\bibnamefont
  {Hautier}}, \bibinfo {author} {\bibfnamefont {C.~C.}\ \bibnamefont
  {Fischer}}, \bibinfo {author} {\bibfnamefont {A.}~\bibnamefont {Jain}},
  \bibinfo {author} {\bibfnamefont {T.}~\bibnamefont {Mueller}}, \ and\
  \bibinfo {author} {\bibfnamefont {G.}~\bibnamefont {Ceder}},\ }\href@noop {}
  {\bibfield  {journal} {\bibinfo  {journal} {Chem. Mater.}\ }\textbf {\bibinfo
  {volume} {22}},\ \bibinfo {pages} {3762} (\bibinfo {year}
  {2010}{\natexlab{a}})}\BibitemShut {NoStop}%
\bibitem [{\citenamefont {Hautier}\ \emph
  {et~al.}(2010{\natexlab{b}})\citenamefont {Hautier}, \citenamefont {Fischer},
  \citenamefont {Ehrlacher}, \citenamefont {Jain},\ and\ \citenamefont
  {Ceder}}]{hautier2010data}%
  \BibitemOpen
  \bibfield  {author} {\bibinfo {author} {\bibfnamefont {G.}~\bibnamefont
  {Hautier}}, \bibinfo {author} {\bibfnamefont {C.}~\bibnamefont {Fischer}},
  \bibinfo {author} {\bibfnamefont {V.}~\bibnamefont {Ehrlacher}}, \bibinfo
  {author} {\bibfnamefont {A.}~\bibnamefont {Jain}}, \ and\ \bibinfo {author}
  {\bibfnamefont {G.}~\bibnamefont {Ceder}},\ }\href@noop {} {\bibfield
  {journal} {\bibinfo  {journal} {Inorg. Chem.}\ }\textbf {\bibinfo {volume}
  {50}},\ \bibinfo {pages} {656} (\bibinfo {year}
  {2010}{\natexlab{b}})}\BibitemShut {NoStop}%
\bibitem [{\citenamefont {Seko}\ \emph {et~al.}(2017)\citenamefont {Seko},
  \citenamefont {Hayashi}, \citenamefont {Nakayama}, \citenamefont
  {Takahashi},\ and\ \citenamefont {Tanaka}}]{PhysRevB.95.144110}%
  \BibitemOpen
  \bibfield  {author} {\bibinfo {author} {\bibfnamefont {A.}~\bibnamefont
  {Seko}}, \bibinfo {author} {\bibfnamefont {H.}~\bibnamefont {Hayashi}},
  \bibinfo {author} {\bibfnamefont {K.}~\bibnamefont {Nakayama}}, \bibinfo
  {author} {\bibfnamefont {A.}~\bibnamefont {Takahashi}}, \ and\ \bibinfo
  {author} {\bibfnamefont {I.}~\bibnamefont {Tanaka}},\ }\href {\doibase
  10.1103/PhysRevB.95.144110} {\bibfield  {journal} {\bibinfo  {journal} {Phys.
  Rev. B}\ }\textbf {\bibinfo {volume} {95}},\ \bibinfo {pages} {144110}
  (\bibinfo {year} {2017})}\BibitemShut {NoStop}%
\bibitem [{\citenamefont {Seko}\ \emph {et~al.}()\citenamefont {Seko},
  \citenamefont {Hayashi},\ and\ \citenamefont {Tanaka}}]{seko-recommender1}%
  \BibitemOpen
  \bibfield  {author} {\bibinfo {author} {\bibfnamefont {A.}~\bibnamefont
  {Seko}}, \bibinfo {author} {\bibfnamefont {H.}~\bibnamefont {Hayashi}}, \
  and\ \bibinfo {author} {\bibfnamefont {I.}~\bibnamefont {Tanaka}},\
  }\href@noop {} {}\bibinfo {note} {Submitted.}\BibitemShut {Stop}%
\bibitem [{\citenamefont {Resnick}\ and\ \citenamefont
  {Varian}(1997)}]{resnick1997recommender}%
  \BibitemOpen
  \bibfield  {author} {\bibinfo {author} {\bibfnamefont {P.}~\bibnamefont
  {Resnick}}\ and\ \bibinfo {author} {\bibfnamefont {H.~R.}\ \bibnamefont
  {Varian}},\ }\href@noop {} {\bibfield  {journal} {\bibinfo  {journal}
  {Commun. ACM}\ }\textbf {\bibinfo {volume} {40}},\ \bibinfo {pages} {56}
  (\bibinfo {year} {1997})}\BibitemShut {NoStop}%
\bibitem [{\citenamefont {Aggarwal}(2016)}]{aggarwal2016recommender}%
  \BibitemOpen
  \bibfield  {author} {\bibinfo {author} {\bibfnamefont {C.~C.}\ \bibnamefont
  {Aggarwal}},\ }\href@noop {} {\emph {\bibinfo {title} {Recommender
  systems}}}\ (\bibinfo  {publisher} {Springer},\ \bibinfo {year}
  {2016})\BibitemShut {NoStop}%
\bibitem [{\citenamefont {Sarwar}\ \emph {et~al.}(2000)\citenamefont {Sarwar},
  \citenamefont {Karypis}, \citenamefont {Konstan},\ and\ \citenamefont
  {Riedl}}]{sarwar2000application}%
  \BibitemOpen
  \bibfield  {author} {\bibinfo {author} {\bibfnamefont {B.}~\bibnamefont
  {Sarwar}}, \bibinfo {author} {\bibfnamefont {G.}~\bibnamefont {Karypis}},
  \bibinfo {author} {\bibfnamefont {J.}~\bibnamefont {Konstan}}, \ and\
  \bibinfo {author} {\bibfnamefont {J.}~\bibnamefont {Riedl}},\ }\href@noop {}
  {\emph {\bibinfo {title} {Application of dimensionality reduction in
  recommender system-a case study}}},\ \bibinfo {type} {Tech. Rep.}\ (\bibinfo
  {institution} {Minnesota Univ Minneapolis Dept of Computer Science},\
  \bibinfo {year} {2000})\BibitemShut {NoStop}%
\bibitem [{\citenamefont {Frolov}\ and\ \citenamefont
  {Oseledets}(2017)}]{frolov2017tensor}%
  \BibitemOpen
  \bibfield  {author} {\bibinfo {author} {\bibfnamefont {E.}~\bibnamefont
  {Frolov}}\ and\ \bibinfo {author} {\bibfnamefont {I.}~\bibnamefont
  {Oseledets}},\ }\href@noop {} {\bibfield  {journal} {\bibinfo  {journal}
  {Wiley Interdiscip. Rev. Data Min. Knowl. Discov.}\ }\textbf {\bibinfo
  {volume} {7}},\ \bibinfo {pages} {e1201} (\bibinfo {year}
  {2017})}\BibitemShut {NoStop}%
\bibitem [{\citenamefont {Symeonidis}\ and\ \citenamefont
  {Zioupos}(2016)}]{symeonidis2016matrix}%
  \BibitemOpen
  \bibfield  {author} {\bibinfo {author} {\bibfnamefont {P.}~\bibnamefont
  {Symeonidis}}\ and\ \bibinfo {author} {\bibfnamefont {A.}~\bibnamefont
  {Zioupos}},\ }\href@noop {} {\emph {\bibinfo {title} {Matrix and Tensor
  Factorization Techniques for Recommender Systems}}}\ (\bibinfo  {publisher}
  {Springer},\ \bibinfo {year} {2016})\BibitemShut {NoStop}%
\bibitem [{\citenamefont {Pedregosa}\ \emph {et~al.}(2011)\citenamefont
  {Pedregosa}, \citenamefont {Varoquaux}, \citenamefont {Gramfort},
  \citenamefont {Michel}, \citenamefont {Thirion}, \citenamefont {Grisel},
  \citenamefont {Blondel}, \citenamefont {Prettenhofer}, \citenamefont {Weiss},
  \citenamefont {Dubourg}, \citenamefont {Vanderplas}, \citenamefont {Passos},
  \citenamefont {Cournapeau}, \citenamefont {Brucher}, \citenamefont {Perrot},\
  and\ \citenamefont {Duchesnay}}]{scikit-learn}%
  \BibitemOpen
  \bibfield  {author} {\bibinfo {author} {\bibfnamefont {F.}~\bibnamefont
  {Pedregosa}}, \bibinfo {author} {\bibfnamefont {G.}~\bibnamefont
  {Varoquaux}}, \bibinfo {author} {\bibfnamefont {A.}~\bibnamefont {Gramfort}},
  \bibinfo {author} {\bibfnamefont {V.}~\bibnamefont {Michel}}, \bibinfo
  {author} {\bibfnamefont {B.}~\bibnamefont {Thirion}}, \bibinfo {author}
  {\bibfnamefont {O.}~\bibnamefont {Grisel}}, \bibinfo {author} {\bibfnamefont
  {M.}~\bibnamefont {Blondel}}, \bibinfo {author} {\bibfnamefont
  {P.}~\bibnamefont {Prettenhofer}}, \bibinfo {author} {\bibfnamefont
  {R.}~\bibnamefont {Weiss}}, \bibinfo {author} {\bibfnamefont
  {V.}~\bibnamefont {Dubourg}}, \bibinfo {author} {\bibfnamefont
  {J.}~\bibnamefont {Vanderplas}}, \bibinfo {author} {\bibfnamefont
  {A.}~\bibnamefont {Passos}}, \bibinfo {author} {\bibfnamefont
  {D.}~\bibnamefont {Cournapeau}}, \bibinfo {author} {\bibfnamefont
  {M.}~\bibnamefont {Brucher}}, \bibinfo {author} {\bibfnamefont
  {M.}~\bibnamefont {Perrot}}, \ and\ \bibinfo {author} {\bibfnamefont
  {E.}~\bibnamefont {Duchesnay}},\ }\href@noop {} {\bibfield  {journal}
  {\bibinfo  {journal} {J. Mach. Learn. Res.}\ }\textbf {\bibinfo {volume}
  {12}},\ \bibinfo {pages} {2825} (\bibinfo {year} {2011})}\BibitemShut
  {NoStop}%
\bibitem [{\citenamefont {Hitchcock}(1927)}]{hitchcock1927expression}%
  \BibitemOpen
  \bibfield  {author} {\bibinfo {author} {\bibfnamefont {F.~L.}\ \bibnamefont
  {Hitchcock}},\ }\href@noop {} {\bibfield  {journal} {\bibinfo  {journal}
  {Stud. Appl. Math.}\ }\textbf {\bibinfo {volume} {6}},\ \bibinfo {pages}
  {164} (\bibinfo {year} {1927})}\BibitemShut {NoStop}%
\bibitem [{\citenamefont {Tucker}(1966)}]{tucker1966some}%
  \BibitemOpen
  \bibfield  {author} {\bibinfo {author} {\bibfnamefont {L.~R.}\ \bibnamefont
  {Tucker}},\ }\href@noop {} {\bibfield  {journal} {\bibinfo  {journal}
  {Psychometrika}\ }\textbf {\bibinfo {volume} {31}},\ \bibinfo {pages} {279}
  (\bibinfo {year} {1966})}\BibitemShut {NoStop}%
\bibitem [{\citenamefont {De~Lathauwer}\ \emph {et~al.}(2000)\citenamefont
  {De~Lathauwer}, \citenamefont {De~Moor},\ and\ \citenamefont
  {Vandewalle}}]{de2000multilinear}%
  \BibitemOpen
  \bibfield  {author} {\bibinfo {author} {\bibfnamefont {L.}~\bibnamefont
  {De~Lathauwer}}, \bibinfo {author} {\bibfnamefont {B.}~\bibnamefont
  {De~Moor}}, \ and\ \bibinfo {author} {\bibfnamefont {J.}~\bibnamefont
  {Vandewalle}},\ }\href@noop {} {\bibfield  {journal} {\bibinfo  {journal}
  {SIAM J. Matrix Anal. Appl.}\ }\textbf {\bibinfo {volume} {21}},\ \bibinfo
  {pages} {1253} (\bibinfo {year} {2000})}\BibitemShut {NoStop}%
\bibitem [{\citenamefont {Nickel}()}]{scikit-tensor}%
  \BibitemOpen
  \bibfield  {author} {\bibinfo {author} {\bibfnamefont {M.}~\bibnamefont
  {Nickel}},\ }\href@noop {} {}\bibinfo {note} {\textsc{scikit-tensor}.
  Available Online. (2013)}\BibitemShut {NoStop}%
\bibitem [{ICD()}]{ICDD}%
  \BibitemOpen
  \href@noop {} {}\bibinfo {note} {ICDD (2010). PDF-4/Organics 2011 (Database),
  edited by Kabekkodu, S., International Centre for Diffraction Data, Newtown
  Square, PA, USA.}\BibitemShut {Stop}%
\bibitem [{Spr()}]{SpringerMaterials}%
  \BibitemOpen
  \href@noop {} {}\bibinfo {note} {SpringerMaterials,
  http://materials.springer.com}\BibitemShut {NoStop}%
\bibitem [{\citenamefont {Bl{\"o}chl}(1994)}]{PAW1}%
  \BibitemOpen
  \bibfield  {author} {\bibinfo {author} {\bibfnamefont {P.~E.}\ \bibnamefont
  {Bl{\"o}chl}},\ }\href@noop {} {\bibfield  {journal} {\bibinfo  {journal}
  {Phys. Rev. B}\ }\textbf {\bibinfo {volume} {50}},\ \bibinfo {pages} {17953}
  (\bibinfo {year} {1994})}\BibitemShut {NoStop}%
\bibitem [{\citenamefont {Kresse}\ and\ \citenamefont {Joubert}(1999)}]{PAW2}%
  \BibitemOpen
  \bibfield  {author} {\bibinfo {author} {\bibfnamefont {G.}~\bibnamefont
  {Kresse}}\ and\ \bibinfo {author} {\bibfnamefont {D.}~\bibnamefont
  {Joubert}},\ }\href@noop {} {\bibfield  {journal} {\bibinfo  {journal} {Phys.
  Rev. B}\ }\textbf {\bibinfo {volume} {59}},\ \bibinfo {pages} {1758}
  (\bibinfo {year} {1999})}\BibitemShut {NoStop}%
\bibitem [{\citenamefont {Perdew}\ \emph {et~al.}(1996)\citenamefont {Perdew},
  \citenamefont {Burke},\ and\ \citenamefont {Ernzerhof}}]{GGA:PBE96}%
  \BibitemOpen
  \bibfield  {author} {\bibinfo {author} {\bibfnamefont {J.~P.}\ \bibnamefont
  {Perdew}}, \bibinfo {author} {\bibfnamefont {K.}~\bibnamefont {Burke}}, \
  and\ \bibinfo {author} {\bibfnamefont {M.}~\bibnamefont {Ernzerhof}},\
  }\href@noop {} {\bibfield  {journal} {\bibinfo  {journal} {Phys. Rev. Lett.}\
  }\textbf {\bibinfo {volume} {77}},\ \bibinfo {pages} {3865} (\bibinfo {year}
  {1996})}\BibitemShut {NoStop}%
\bibitem [{\citenamefont {Kresse}\ and\ \citenamefont {Hafner}(1993)}]{VASP1}%
  \BibitemOpen
  \bibfield  {author} {\bibinfo {author} {\bibfnamefont {G.}~\bibnamefont
  {Kresse}}\ and\ \bibinfo {author} {\bibfnamefont {J.}~\bibnamefont
  {Hafner}},\ }\href@noop {} {\bibfield  {journal} {\bibinfo  {journal} {Phys.
  Rev. B}\ }\textbf {\bibinfo {volume} {47}},\ \bibinfo {pages} {558} (\bibinfo
  {year} {1993})}\BibitemShut {NoStop}%
\bibitem [{\citenamefont {Kresse}\ and\ \citenamefont
  {Furthm{\"u}ller}(1996)}]{VASP2}%
  \BibitemOpen
  \bibfield  {author} {\bibinfo {author} {\bibfnamefont {G.}~\bibnamefont
  {Kresse}}\ and\ \bibinfo {author} {\bibfnamefont {J.}~\bibnamefont
  {Furthm{\"u}ller}},\ }\href@noop {} {\bibfield  {journal} {\bibinfo
  {journal} {Phys. Rev. B}\ }\textbf {\bibinfo {volume} {54}},\ \bibinfo
  {pages} {11169} (\bibinfo {year} {1996})}\BibitemShut {NoStop}%
\end{thebibliography}%

\end{document}